\documentclass[%
 reprint,
superscriptaddress,
 amsmath,amssymb,
 aps,
prl,
]{revtex4-2}

\usepackage{dcolumn}

\usepackage[T1]{fontenc}
\usepackage[utf8]{inputenc}
\usepackage{bbold}
\usepackage{amsmath}
\usepackage{amssymb}
\usepackage{graphicx}
\usepackage{graphicx}
\usepackage{subscript}
\usepackage[unicode=true,pdfusetitle,
bookmarks=true,bookmarksnumbered=false,bookmarksopen=false,
breaklinks=false,pdfborder={0 0 1},backref=false,colorlinks=false]
{hyperref}
\usepackage{xcolor}
\usepackage[title]{appendix}

\newcommand{\lyxmathsym}[1]{\ifmmode\begingroup\def\b@ld{bold}
	\text{\ifx\math@version\b@ld\bfseries\fi#1}\endgroup\else#1\fi}

\begin{document}
\title{Quantum valley pseudospin controlled by strain}

\author{Maur\'icio F. C. Martins Quintela}
\email{mfcmquintela@gmail.com, corresponding author}
\affiliation{Departamento de Qu\'{i}mica, Universidad Aut\'{o}noma de Madrid, E-28049 Madrid, Spain}
\affiliation{Condensed Matter Physics Center (IFIMAC), Universidad Aut\'{o}noma de Madrid, E-28049, Madrid, Spain}
\affiliation{Departamento de F\'isica de la Materia Condensada, Universidad Aut\'{o}noma de Madrid, E-28049 Madrid, Spain}

\author{Miguel S\'a}
\affiliation{Departamento de Qu\'{i}mica, Universidad Aut\'{o}noma de Madrid, E-28049 Madrid, Spain}
\affiliation{Condensed Matter Physics Center (IFIMAC), Universidad Aut\'{o}noma de Madrid, E-28049, Madrid, Spain}
\affiliation{Departamento de F\'isica de la Materia Condensada, Universidad Aut\'{o}noma de Madrid, E-28049 Madrid, Spain}

\author{Alejandro J. Ur\'ia-\'Alvarez}%
\affiliation{Departamento de F\'isica de la Materia Condensada, Universidad Aut\'{o}noma de Madrid, E-28049 Madrid, Spain}

\author{Mikhail Malakhov}%
\affiliation{M.N. Mikheev Institute of Metal Physics of the Ural Branch of the Russian Academy of Sciences, S. Kovalevskaya str. 18, 620108 Yekaterinburg, Russia}

\author{Giovanni Cistaro}%
\affiliation{Departamento de Qu\'{i}mica, Universidad Aut\'{o}noma de Madrid, E-28049 Madrid, Spain}
\affiliation{Theory and Simulation of Materials (THEOS), \'Ecole Polytechnique F\'ed\'erale de Lausanne (EPFL), CH-1015 Lausanne, Switzerland}

\author{Jorge Quereda}
\affiliation{Instituto de Ciencia de Materiales de Madrid (ICMM-CSIC), Madrid, E-28049 Spain}

\author{Juan J. Palacios}
\affiliation{Condensed Matter Physics Center (IFIMAC), Universidad Aut\'{o}noma de Madrid, E-28049, Madrid, Spain}
\affiliation{Departamento de F\'isica de la Materia Condensada, Universidad Aut\'{o}noma de Madrid, E-28049 Madrid, Spain}
\affiliation{Instituto Nicol\'as Cabrera (INC), Universidad Aut\'onoma de Madrid, E-28049 Madrid, Spain}

\author{Antonio Pic\'on}
\email{antonio.picon@csic.es, corresponding author}
\affiliation{Departamento de Qu\'{i}mica, Universidad Aut\'{o}noma de Madrid, E-28049 Madrid, Spain}
\affiliation{Condensed Matter Physics Center (IFIMAC), Universidad Aut\'{o}noma de Madrid, E-28049, Madrid, Spain}
\affiliation{Instituto de Ciencia de Materiales de Madrid (ICMM-CSIC), Madrid, E-28049 Spain}

\begin{abstract}

Valleytronics, as an alternative to traditional electronics or spintronics, is based on the encoding of quantum information in pseudospin valley quantum numbers, rather than in charge or spin states. A key ingredient is the (optical) manipulation of valley states before loss of coherence, which can be as fast as 100 femtoseconds. Previous works have shown the possibility of valley state manipulation using external fields. Here we propose uniaxial strain as a more flexible and robust scheme to manipulate the valley state through the breaking of the crystal symmetry and the concomitant lifting of the degeneracy of the $1s$ exciton energy. Our theory is corroborated by state-of-the-art numerical simulations in monolayer hBN and shows the possibility to control valley pseudospin at the attosecond time scale.


\end{abstract}

\date{\today}
\maketitle


The breaking of symmetries usually gives rise to interesting phenomena such as spin-polarization effects, nonlinear effects, superconductivity, and valleytronics, to just name a few \cite{du2021}, with a wide variety of technology applications and advances in photonics and optoelectronics. The term valleytronics refers to the possibility of encoding information through the crystal momentum of the electrons, specifically in the so-called valley pseudospin, which refers to the momentum at (optically) relevant time-reversed points in the Brillouin zone. The unique properties of two-dimensional (2D) materials are essential to deploy the potential of valleytronics \cite{yao2012}. In monolayer materials such as hBN and transition metal dichalcogenides (TMDC), inversion symmetry breaking gives rise to a nontrivial Berry connection around $K$ and $K'$ points in reciprocal space. This allows for selection rules that permit to selectively excite carriers at the $K$ or $K'$ valleys, by using circularly polarized pulses \cite{yao2012}. This selectivity can be used to encode information, opening a promising framework for electronics and quantum information. Furthermore, it is based on an all-optical scheme, which enables to manipulate the information at the ultimate speed of electrons, \emph{i.e.} at the attosecond timescale \cite{vitale2018,freudenstein2022}.

\begin{figure*}
	\centering
	\begin{centering}
		\hspace{-0.5cm}\includegraphics[scale=0.65]{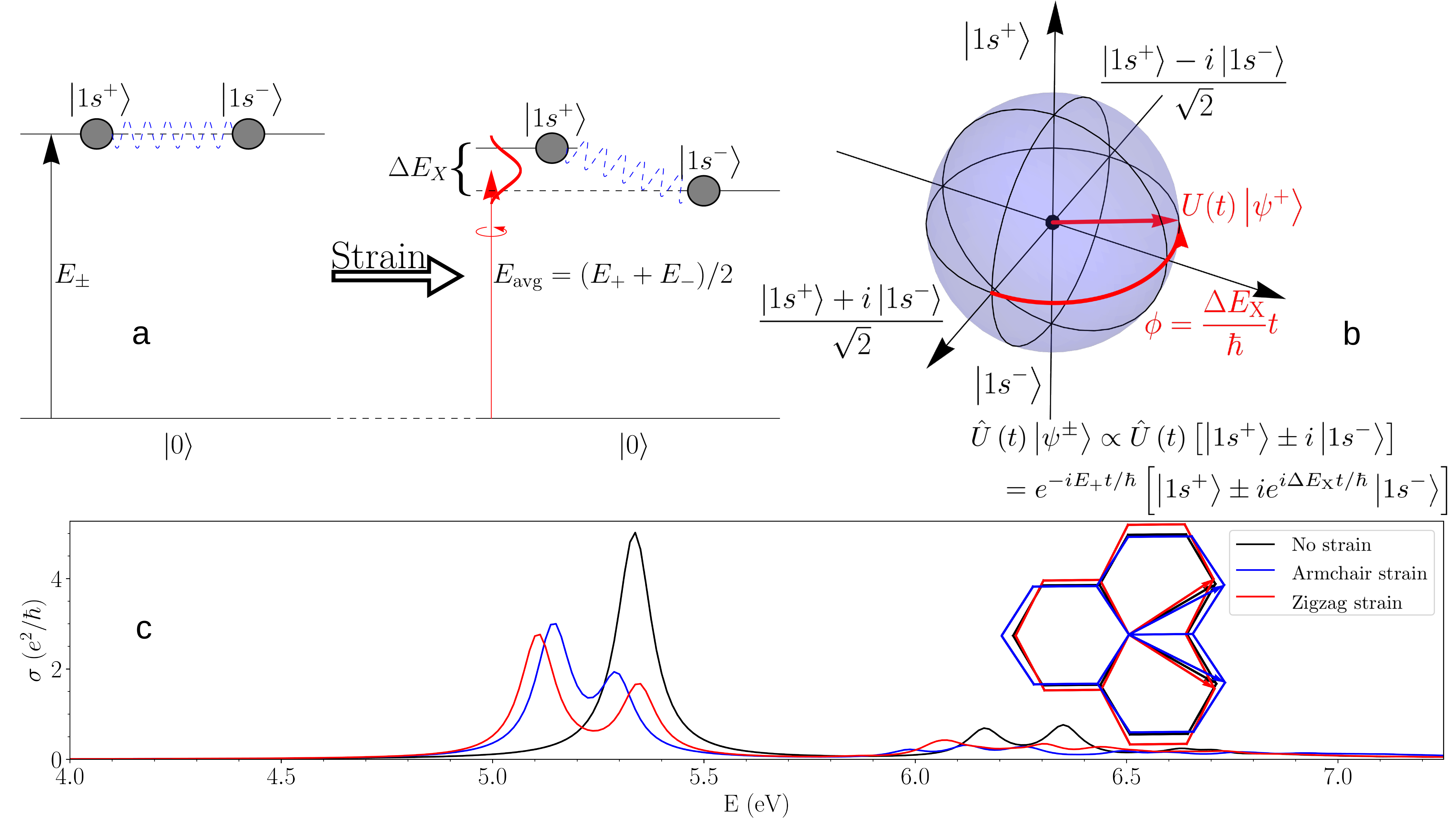}
		\par\end{centering}
	\caption{\textbf{a)} Scheme of the energy degeneracy lifting of $1s$ exciton states due to strain, together with the pulse designed to excite the two relevant states.  
    \textbf{b)} View of the rotations of the excitonic quantum states on the pseudospin Bloch sphere induced by strain. \textbf{c)} Circularly polarized optical conductivity for unstrained system (black line) as well as for the system under armchair (blue line) and zigzag strain (red line), as obtained through XATU \cite{URIAALVAREZ2024109001}. Inset represents the lattice deformation for each corresponding strain.
}\label{fig:fig1}
\end{figure*}

 In 2D materials, due to their low dimensionality and reduced screening, the optical response is dominated by excitons. Excitons can be considered quasi-particles composed by an electron-hole pair bound via Coulomb interaction. In particular, the degenerate 1$s$ excitonic state dominates the absorption spectrum. Previous works demonstrated rotations on the pseudospin Bloch sphere of the excitonic quantum states, see Fig. \ref{fig:fig1}b, by using magnetic or electric external fields \cite{wang2016,ye2017,sie2017}. 

Current technology allows to break and control the symmetry of 2D materials by relying on external fields, layer stacking, or strain \cite{du2021}. In particular, the potential of strain engineering to tailor the valley-related behaviour of carriers in TMDCs has been vastly explored in the literature \cite{zhao2021}. Despite the relevance of the $1s$ exciton peak in the optical response, strain effects on excitons have been barely investigated. In contrast to previous studies, in which an external magnetic field \cite{wang2016} or a laser pulse field \cite{ye2017,sie2017} is used to modify the exciton energies and produce a rotation in the Bloch sphere, here we exploit strain. We prove that strain, by breaking the spatial symmetry, lifts the $1s$ exciton degeneracy and consequently the excitation with a circularly polarized pulse induces a valley rotation. This has two main advantages, i) the degeneracy can be permanently lifted and ii) the orientation and the strength of the strain offer an additional control knob without the energy cost of an external field, which is more attractive for technology purposes. We show that the speed of the full valley pseudospin rotation could reach the few-femtosecond time scale with realistic strains, being comparable or superior to reported schemes with external fields \cite{wang2016,ye2017,sie2017,zhao2021,schmidt2016,gucci2024}. 
 

Detecting the valley state is a challenging task due to its short coherence lifetime, which is reported in TMDCs to lie between $98-520\,\mathrm{fs}$ \cite{wang2016,gupta2023}. It is then desired a fast valley-pseudospin manipulation and therefore also a detection scheme to track the state changes. Previous detection schemes are based on photoluminescence \cite{ye2017}, two-dimensional coherent spectroscopy \cite{hao2016}, second-harmonic generation \cite{herrmann2024}, and Faraday rotation \cite{gucci2024}. Here, due to our fast few-femtosecond time scales to manipulate the valley pseudospin rotation, we explore the possibility to follow the state changes via attosecond transient absorption spectroscopy (ATAS). The recent generation of attosecond ($10^{-18}\,\mathrm{s}$) pulses \cite{RevModPhys.96.030503} allowed us to perform ultrafast (pump-probe) experiments to track dynamics at the ultimate speed of electron motion. ATAS has been successfully applied to track exciton dynamics \cite{Moulet2017,Lucchini2021}, and it is sensitive to the coherent laser-induced dynamics \cite{PhysRevResearch.3.013144,Malakhov2024} and to the topological phase \cite{mosquera2024}. In this context, we provide the interpretation of the spectral features of a possible ATAS study that would enable us to retrieve the valley state changes. 

In this manuscript we focus on monolayer hBN, which enables us to simplify the system and introduce the main concept, but the theory can be easily extended to TMDCs. Our theoretical results are supported by state-of-the-art real-time simulations performed with the EDUS code \cite{doi:10.1021/acs.jctc.2c00674}, which accounts both for light-matter and electron-electron interactions.


\begin{figure*}
	\centering
	\begin{centering}
        \includegraphics[scale=0.66]{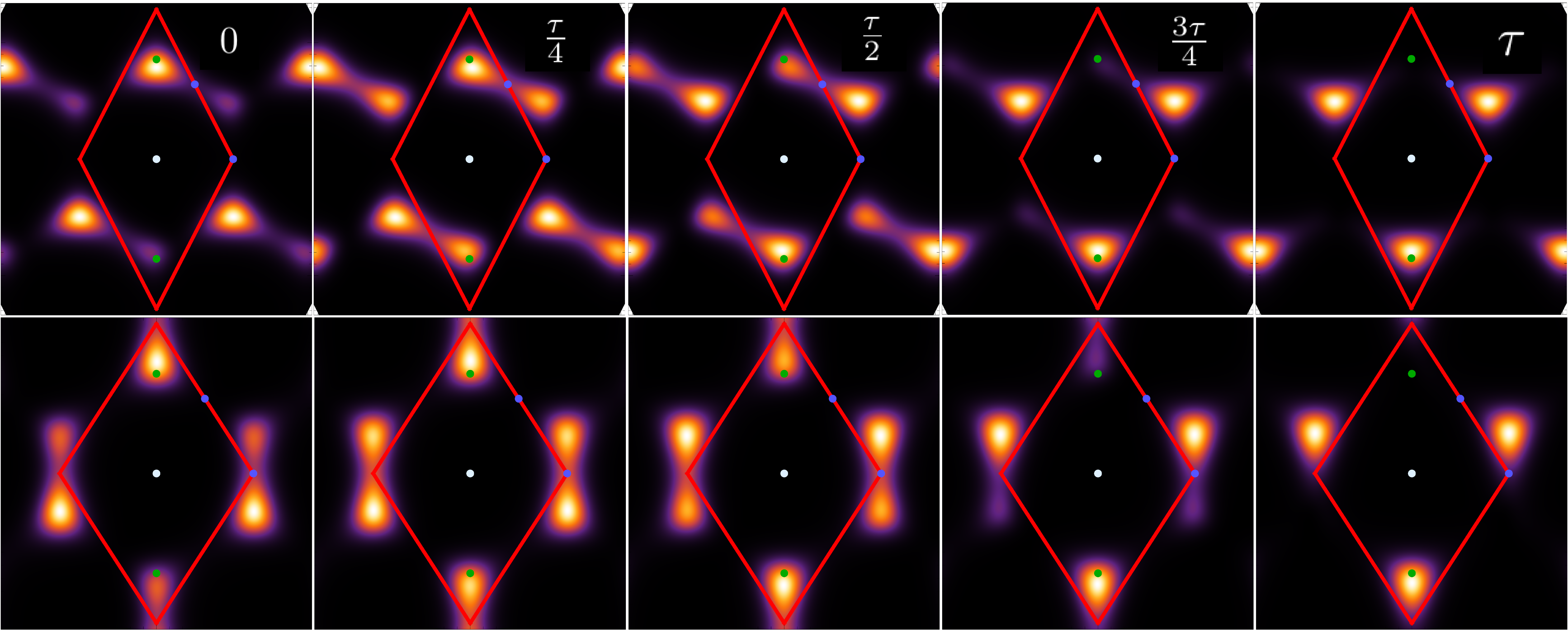}
    \par\end{centering}
	\caption{Time evolution (left--to--right) of the exciton wavefunction $\rho_{cv}({\bf k})$, from $t\approx t_f$ to $t\approx t_f + \tau$, right after excitation under $\epsilon=0.1$ strain, computed via EDUS \cite{doi:10.1021/acs.jctc.2c00674}. Top row: armchair ($\theta_{s}=0$) strain. Bottom row: zigzag ($\theta_{s}=\pi/2$) strain. Fifth panel in each row corresponds to $t\approx t_f + \tau$, where the $\left|\psi^{+}\right>$ state has fully transitioned into the $\left|\psi^{-}\right>$ state. Red outline denotes the first Brillouin zone, with points green, white, and blue representing the ${\bf K}/{\bf K}'$, $\mathbf{\Gamma}$, and $\mathbf{M}/\mathbf{M}^\prime$ points, respectively.
 }\label{fig:fig2}
\end{figure*}

A circularly polarized laser pulse can selectively excite a valley. This is dictated by the valley-pseudospin selection rule, which is the essence for valleytronics in hBN and TMDCs \cite{PhysRevB.77.235406}. For $1s$ exciton states, this selection rule is also inherited \cite{PhysRevLett.128.047402}. For an unstrained system, see Fig. \ref{fig:fig1}a, circularly polarized light excites a linear combination of the two degenerate $1s$ states as
\begin{equation}
\left|\psi^{\pm}\right>=\frac{\left|1s^{+}\right>\pm i\left|1s^{-}\right>}{\sqrt{2}},
\label{eq:circularexcitons}
\end{equation}
that leads to a valley localization. The $\pm$ superscript corresponds to positive/negative circular polarization. When the strain is introduced, the anisotropic optical response would change the excitation from circularly polarized light, which can be expressed as a sum of the linear conductivity for linear polarization in the x- and y- direction, see Supplemental Material. We calculate the optical conductivity for unstrained and strained monolayer hBN, see Fig. \ref{fig:fig1}c, using the standard Bethe-Salpeter equation (BSE). In the BSE, a superposition of electron-hole pairs in the reciprocal space is used as an \emph{ansatz} to solve the Hamiltonian with electron-hole interactions \cite{HIRATA1999291,doi:10.1021/cr0505627,PhysRevB.89.085310}. We consider a two-band tight binding model to describe the electronic structure of hBN, see more details in the Supplemental Material, and we employ the XATU code to solve the BSE \cite{URIAALVAREZ2024109001}. In the unstrained system, the first dominant peak in the optical conductivity around $5.4\,\mathrm{eV}$ corresponds to the two degenerate $1s$ excitons, which are located below the bandgap (in our case at $7.25\,\mathrm{eV}$). In the strained system, the energy degeneracy of the $1s$ excitons is broken and the absorption peaks split into two. We consider strain strengths ($\epsilon$) of $10\%$. The optical and electronic properties of 2D materials can be tailored to a large extent via external mechanical deformation. While bulk crystalline materials can rarely sustain mechanical strains beyond $1\%$ without breaking, 2D crystals can withstand extreme deformations, reaching strain limits of up to $25\%$ for graphene, $20\%$ for hBN and $11\%$ for $\mathrm{MoS}_2$, to name a few \cite{Peng2020}. Furthermore, the orientation of the uniaxial strain ($\theta_s$) also offers an additional knob to split the $1s$ excitons peaks. We observe in Fig. \ref{fig:fig1}c how the zigzag ($\theta_s=\pi/2$) strain provides a larger energy splitting than the one induced by the armchair ($\theta_s=0$) strain.

\begin{figure*}
	\centering
	\begin{centering}
		\includegraphics[scale=0.66]{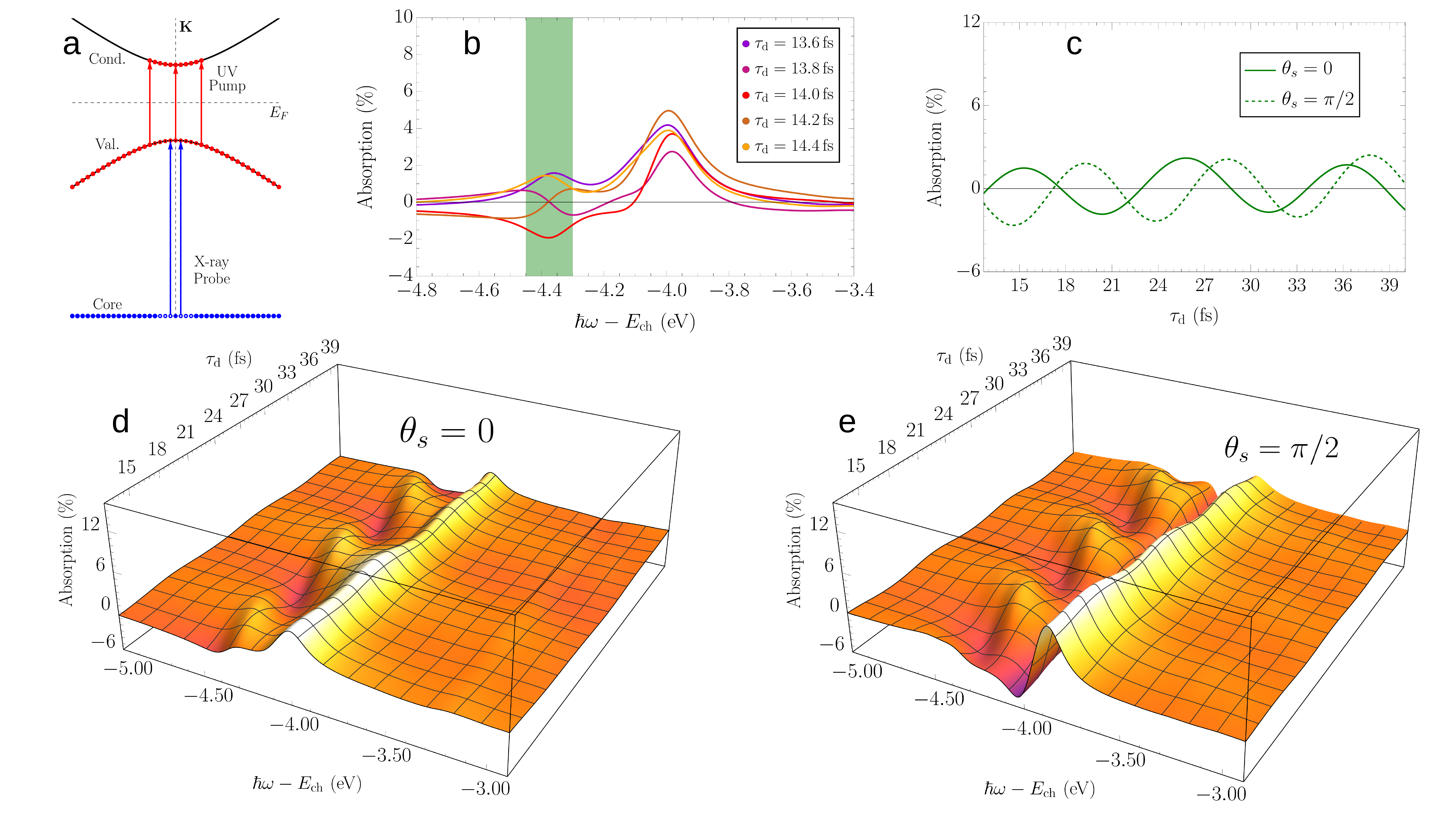}
    \par\end{centering}
	\caption{
    \textbf{a)} Illustration of the relevant transitions involved in the pump-probe signal.  
    \textbf{b)} Transient absorption spectrum in the energy range corresponding to transitions from core to valence states. 
    Fast oscillations are observed with a period $T_{\tau_+}\approx0.8\,\mathrm{fs}$, consistent with the probe pulse capturing the global phase of an exciton coherent state. Data for hBN strained along the armchair direction ($\epsilon=0.1$, $\theta_s=0$). The integrated area around the peak at -4.4 eV is highlighted in green. 
    \textbf{c)} Oscillations in time delay $\tau_\mathrm{d}$ for the peak around -4.4 eV for both armchair (solid lines) and zigzag (dashed lines) strain directions.
    \textbf{d)} and \textbf{e)} Transient absorption spectrum in hBN strained along the armchair ($\epsilon=0.1$, $\theta_s=0$) and zigzag ($\epsilon=0.1$, $\theta_s=\pi/2$) direction, respectively. 
    }\label{fig:fig3}
\end{figure*}

It seems reasonable to use an ultrashort (broad bandwidth) pulse to excite both $1s$ states
and produce a superposition as in Eq. \ref{eq:circularexcitons}. However, due to the lifted degeneracy we expect the time evolution of the superposition to be
\begin{align}
	&\hat{U}\left(t\right)\left|\psi^{\pm}\right> \propto\hat{U}\left(t\right)\left[\left|1s^{+}\right>\pm i\left|1s^{-}\right>\right]\nonumber \\
	&\quad =e^{-i E_{+}t/\hbar}\left[ \left|1s^{+}\right> \pm ie^{i\Delta E_{\mathrm{X}}t/\hbar}\left|1s^{-}\right>\right] ,\label{eq:psi_time-evolve}
\end{align}
where $\Delta E_{\mathrm{X}}=E_{+}-E_{-}$ is the energy splitting between the eigenvalues of the states $\left|1s^{+/-}\right>$. While the global $e^{-iE_{+}t/\hbar}$ factor simply introduces a global phase to the state as it evolves in time, the $e^{i\Delta E_{\mathrm{X}} t/\hbar}$ factor multiplying $\left|1s^{-}\right>$ gives rise to the rotation of the valley state, which is represented by a rotation around the $z-$axis in the Bloch sphere of Fig. \ref{fig:fig1}b. The presence of this complex exponential means that the phase of the $\left|1s^{-}\right>$ state changes in time and, more specifically, after a time
\begin{equation}
\tau=\frac{\pi\hbar}{\Delta E_{\mathrm{X}}},\label{eq:valley_migration_time}
\end{equation}
the state $\left|\psi^{\pm}\right>$ evolves to
\begin{align}
\hat{U}\left(\tau\right)\left|\psi^{\pm}\right> & = e^{-i\frac{E_{+}}{\Delta E_{\mathrm{X}}}\pi}\left|\psi^{\mp}\right>.
\label{eq:psipm}
\end{align}
Hence, the time $\tau$ defines a ``valley migration time'' between two states $\left|\psi^{+/-}\right>$, mostly localized in the $K/K^\prime$ valleys, respectively. 
In order to prove the strain-tunable valley rotation, we perform real-time simulations in strained hBN with the EDUS code \cite{doi:10.1021/acs.jctc.2c00674,PhysRevResearch.3.013144,mosquera2024,Malakhov2024}, which consists in evolving the one-electron reduced density matrix $\rho_{ij}({\bf k})$ in the reciprocal space. We consider a circularly polarized pulse with photon frequency centered at $E_{\mathrm{avg}} = (E_{+}+E_{-})/2$ between the two $1s$ exciton peaks in order to produce the $\left|\psi^{\pm}\right>$ state given by Eq. (\ref{eq:circularexcitons}). The pulse must be short enough to excite both $1s$ exciton peaks. For a strain strength $\epsilon=0.1$, the pulse has a total time duration of approximately $12\,\mathrm{fs}$. See more details about the pump pulse parameters in the Supplemental Material. The evolution of the exciton wavefunction given by the $\rho_{cv}({\bf k})$ term, where the index $c$ ($v$) stands for the conduction (valence) band, is represented in Fig. \ref{fig:fig2}. We observe the valley rotation for the zigzag and armchair strain directions. The panels in each row are plotted at specific time steps $\delta t=n\frac{\tau}{4}$, with the peak localized in the $K$ valley moving from left to right, transitioning into a $K^\prime$ peak at $t=\tau$. Furthermore, these plots also clearly showcase the effects of the deformation of the Brillouin zone on the valley migration. The valley migration time $\tau$ is approximately $12.5\,\mathrm{fs}$ for $\epsilon=0.1$. This demonstrates the control of quantum rotations at the attosecond time scale by using uniaxial strain. 


The exciton quantum state oscillation between valleys relies on coherence and therefore it should be observed before the dynamics couples to other degrees of freedom. Current technology permits the track of exciton dynamics on the attosecond scale via transient absorption or reflectivity spectroscopy \cite{Moulet2017,Lucchini2021}. To simulate an attosecond experiment with the EDUS code, we need to model a pump-probe scheme \cite{Picon_2019,Buades2021}. The first pump pulse excites the material and induces the above mentioned exciton oscillation, while a second probe pulse is absorbed in a later stage, with a certain time delay that is controlled. The probe pulse has photon energies around $410\,\mathrm{eV}$ in order to excite core transitions from the N-$1s$-orbital band to the valence or conduction band. The probe pulse is short enough to have a bandwidth to cover the whole valence and conduction band in our hBN model. We include therefore the N-$1s$-orbital band in our tight-binding model as in Ref. \cite{Malakhov2024}. For the simulations we consider a range of time delays $\Delta\tau_{\mathrm{d}}\in\left[0,\,28\,\mathrm{fs}\right]$ after the pump pulse. 


The pump-probe signal is strong when the probe excites core electrons into the valence holes produced by the pump pulse, see scheme in Fig. \ref{fig:fig3}a. For that particular energy window, we observe strong oscillations in the absorption with respect to the time delay, see Fig. \ref{fig:fig3}b. The energy scale refers to the energy from the N-$1s$ band to the Fermi level, \emph{i.e.} $E_{\mathrm{ch}}=409.9\,\mathrm{eV}$, and the intensity is normalized with respect to the maximum peak, which corresponds to the core-to-conduction excitation. From the way the exciton superposition is created following Eq. (\ref{eq:psi_time-evolve}), at least two distinct periodicities are found in the pump--probe absorption signal, see further details in Supplemental Material. One slow oscillation associated with the valley migration, $T_\tau = \frac{2\pi\hbar}{\Delta E_{\mathrm{X}}}\approx\,12.5\,\mathrm{fs}$, and one fast oscillation associated with the global phase of the wave function at the instant of the probe arrival, $T_{\tau_+} = \frac{2\pi\hbar}{E_+}\approx\,0.8\,\mathrm{fs}$. The fast oscillations are clearly observed in Fig. \ref{fig:fig3}b by taking snapshots with small steps of time delays. By filtering out the fast oscillations, the remaining signal shows the period of the exciton valley oscillations, see Fig. \ref{fig:fig3}c in which the intensity of the peak around $-4.4\,\mathrm{eV}$ is represented as a function of the time delay. We show the whole transient absorption spectrum in Fig. \ref{fig:fig3}d and e for armchair and zigzag strain, respectively. The numerical simulations show the sensitivity of ATAS to follow the valence holes at the N site during the valley state rotation.


In conclusion, we propose a scheme to manipulate the valley pseudospin via uniaxial strain. This is supported by numerical real-time simulations in hBN, which show a perfect valley rotation whose speed is determined by the energy split of the $1s$ excitons, which depends on both the strength and the orientation of the strain. Furthermore, we also demonstrate the feasibility to characterize the valley rotation by using attosecond absorption spectroscopy. It is worth noting that, while here we focused our discussion on hBN, our approach can be easily extended to a wide variety of 2D materials. This work opens novel strategies to harness energy-efficient schemes for valleytronics based on mechanical distortions of the crystal, which can be combined with stacking schemes to increase the valley coherence time \cite{gupta2023}. 




\section*{Acknowledgements}
The authors acknowledge financial support to the Spanish Ministry of Science, Innovation and Universities \& the State Research Agency through grants PID2021-126560NB-I00, CNS2022-135803, TED2021-131323B-I00, and PID2022-141712NB-C21 (MCIU/AEI/FEDER, UE), and the "Mar\'ia de Maeztu" Programme for Units of Excellence in R\&D   (CEX2023-001316-M), the
Comunidad de Madrid and the Spanish State through the Recovery, Transformation and Resilience Plan [Materiales Disruptivos Bidimensionales (2D), MAD2D-CM-UAM7 Materiales Avanzados], the European Union through the Next Generation EU funds, the Generalitat Valenciana through the Program Prometeo (2021/017). M.M. thanks the Ministry of Science and Higher Education of the Russian Federation for supporting theoretical calculations through funding the Institute of Metal Physics. We also acknowledge computer resources and assistance provided by Centro de Computaci\'on Cient\'ifica de la Universidad Aut\'onoma de Madrid and RES resources (FI-2024-3-0010, FI-2024-3-0011, FI-2024-2-0034, FI-2023-2-0012, FI-2022-3-0022, FI-2022-1-0031).


\bibliographystyle{ieeetr}
\bibliography{strain-tuning}
\end{document}


\title{Supplementary Material:\linebreak{} Quantum valley pseudospin controlled by strain}

\author{Maur\'icio F. C. Martins Quintela}
\email{mfcmquintela@gmail.com, corresponding author}
\affiliation{Departamento de Qu\'{i}mica, Universidad Aut\'{o}noma de Madrid, 28049 Madrid, Spain}
\affiliation{Condensed Matter Physics Center (IFIMAC), Universidad Aut\'{o}noma de Madrid, 28049, Madrid, Spain}
\affiliation{Departamento de F\'isica de la Materia Condensada, Universidad Aut\'{o}noma de Madrid, E-28049 Madrid, Spain}

\author{Miguel S\'a}
\affiliation{Departamento de Qu\'{i}mica, Universidad Aut\'{o}noma de Madrid, 28049 Madrid, Spain}
\affiliation{Condensed Matter Physics Center (IFIMAC), Universidad Aut\'{o}noma de Madrid, 28049, Madrid, Spain}
\affiliation{Departamento de F\'isica de la Materia Condensada, Universidad Aut\'{o}noma de Madrid, E-28049 Madrid, Spain}

\author{Alejandro J. Ur\'ia-\'Alvarez}%
\affiliation{Departamento de F\'isica de la Materia Condensada, Universidad Aut\'{o}noma de Madrid, E-28049 Madrid, Spain}

\author{Mikhail Malakhov}%
\affiliation{M.N. Mikheev Institute of Metal Physics of the Ural Branch of the Russian Academy of Sciences, S. Kovalevskaya str. 18, 620108 Yekaterinburg, Russia}

\author{Giovanni Cistaro}%
\affiliation{Departamento de Qu\'{i}mica, Universidad Aut\'{o}noma de Madrid, 28049 Madrid, Spain}
\affiliation{Theory and Simulation of Materials (THEOS), \'Ecole Polytechnique F\'ed\'erale de Lausanne (EPFL), CH-1015 Lausanne, Switzerland}

\author{Jorge Quereda}
\affiliation{Instituto de Ciencia de Materiales de Madrid (ICMM-CSIC), Madrid, 28049 Spain}

\author{Juan J. Palacios}
\affiliation{Condensed Matter Physics Center (IFIMAC), Universidad Aut\'{o}noma de Madrid, 28049, Madrid, Spain}
\affiliation{Departamento de F\'isica de la Materia Condensada, Universidad Aut\'{o}noma de Madrid, E-28049 Madrid, Spain}
\affiliation{Instituto Nicol\'as Cabrera (INC), Universidad Aut\'onoma de Madrid, E-28049 Madrid, Spain}

\author{Antonio Pic\'on}
\email{antonio.picon@csic.es, corresponding author}
\affiliation{Departamento de Qu\'{i}mica, Universidad Aut\'{o}noma de Madrid, 28049 Madrid, Spain}
\affiliation{Condensed Matter Physics Center (IFIMAC), Universidad Aut\'{o}noma de Madrid, 28049, Madrid, Spain}
\affiliation{Instituto de Ciencia de Materiales de Madrid (ICMM-CSIC), Madrid, 28049 Spain}
\maketitle

\tableofcontents

\section*{SM1: Lattice structure of hBN}
\label{sec:appendixA}

To study the effect of deformation on excitons in hBN, we first study the effects on the unit cell, the vectors joining nearest neighbors, and the Bravais vectors. The vectors $\boldsymbol{\delta}_{i}$ are defined as 
\begin{align*}
\boldsymbol{\delta}_{1} & =a_0\left[\begin{array}{c}
1\\
0
\end{array}\right] & \boldsymbol{\delta}_{2} & =a_0\left[\begin{array}{c}
-\frac{1}{2}\\
\frac{\sqrt{3}}{2}
\end{array}\right] & \boldsymbol{\delta}_{3} & =a_0\left[\begin{array}{c}
-\frac{1}{2}\\
-\frac{\sqrt{3}}{2}
\end{array}\right],
\end{align*}
with the unit cell being aligned with $\boldsymbol{\delta}_{1}$ and
the Bravais vectors read
\begin{align*}
\mathbf{a}_{1} & =\boldsymbol{\delta}_{1}-\boldsymbol{\delta}_{3}=a_0\left[\begin{array}{c}
\frac{3}{2}\\
\frac{\sqrt{3}}{2}
\end{array}\right] &
 \mathbf{a}_{2} & =\boldsymbol{\delta}_{1}-\boldsymbol{\delta}_{2}=a_0\left[\begin{array}{c}
\frac{3}{2}\\
-\frac{\sqrt{3}}{2}
\end{array}\right].
\end{align*}
This lattice structure is then portrayed 
in Fig. 1a of the main text. As zigzag strain is not parallel to any of the three $\boldsymbol{\delta}_{i}$ vectors defined above, its effects will be qualitatively distinct from armchair strain. This thus leads to a qualitatively distinct evolution of the hopping parameters
when compared to armchair strain. Consequently, the dispersion energy bands are affected, see more details in the SM, and results in different exciton properties, as discussed in the main text.

Knowing the Bravais vectors, the primitive vectors in reciprocal space read 
\begin{align*}
	\mathbf{b}_{1} & =\frac{2\pi}{\sqrt{3}a_0}\left[\begin{array}{c}
		\frac{1}{\sqrt{3}}\\
		1
	\end{array}\right]&
	\mathbf{b}_{2} & =\frac{2\pi}{\sqrt{3}a_0}\left[\begin{array}{c}
		\frac{1}{\sqrt{3}}\\
		-1
	\end{array}\right],
\end{align*}
which means that the high--symmetry Dirac points are given by 
\begin{align*}
	\mathbf{K} & =\frac{\mathbf{b}_{1}-\mathbf{b}_{2}}{3}=\frac{4\pi}{3\sqrt{3}a_0}\left[\begin{array}{c}
		0\\
		1
	\end{array}\right]&
	\mathbf{K}^{\prime} & =\frac{\mathbf{b}_{2}-\mathbf{b}_{1}}{3}=\frac{4\pi}{3\sqrt{3}a_0}\left[\begin{array}{c}
		0\\
		-1
	\end{array}\right].
\end{align*}

\section*{SM2: Strained Dirac points}
\label{sec:appendixB}

To introduce strain in the hBN lattice, we must first define how uniaxial deformation transforms generic two--dimensional vectors. From the generalized Hooke's law \cite{Silberschmidt2022-hj}, the strain tensor for uniaxial deformation in the lattice coordinate system reads \cite{PhysRevB.80.045401,URIAALVAREZ2024109001}
\begin{align}
\boldsymbol{\epsilon}&=\epsilon\left[\begin{array}{cc}
\cos^{2}\theta_{s}-\sigma\sin^{2}\theta_{s} & \left(1+\sigma\right)\cos\theta_{s}\sin\theta_{s}\\
\left(1+\sigma\right)\cos\theta_{s}\sin\theta_{s} & \sin^{2}\theta_{s}-\sigma\cos^{2}\theta_{s}
\end{array}\right],\label{eq:generic_strain}
\end{align}
where $\theta_{s}$ is the direction of the strain, measured (w.l.o.g.) from the $x-$axis, $\epsilon$ is the strain strength, and $\sigma=0.211$ is the Poisson ratio in hBN \cite{PENG201211,Han_2014,Falin2017}. While the strain effects on the lattice vectors lead to rather unwieldy expressions due to the many cross-terms present in equation \ref{eq:generic_strain}, the strained Dirac points $\mathbf{K}/\mathbf{K}^{\prime}$ can be written simply as
\begin{align*}
	\tilde{\mathbf{K}} & =\frac{2\pi}{3\sqrt{3}a_0}\left[\begin{array}{c}
		-\frac{\epsilon\left(1+\sigma\right)\sin2\theta_s}{\left(1+\epsilon\right)\left(1-\epsilon\sigma\right)}\\
		\\
		\frac{2+\epsilon\left(1-\sigma\right)+\epsilon\left(1+\sigma\right)\cos2\theta_{s}}{\left(1+\epsilon\right)\left(1-\epsilon\sigma\right)}
	\end{array}\right] &
	\tilde{\mathbf{K}^{\prime}} & =\frac{2\pi}{3\sqrt{3}a_0}\left[\begin{array}{c}
		\frac{\epsilon\left(1+\sigma\right)\sin2\theta_s}{\left(1+\epsilon\right)\left(1-\epsilon\sigma\right)}\\
		\\
		-\frac{2+\epsilon\left(1-\sigma\right)+\epsilon\left(1+\sigma\right)\cos2\theta_{s}}{\left(1+\epsilon\right)\left(1-\epsilon\sigma\right)}
	\end{array}\right].
\end{align*}


\section*{SM3: Excitonic conductivity}
\label{sec:appendixC}

The generic component of the linear conductivity is given by \cite{pedersen_intraband_2015,taghizadeh_nonlinear_2019,PhysRevB.107.235416}
\begin{eqnarray}
	\frac{\sigma_{\alpha\beta}\left(\omega\right)}{\sigma_{0}}=\frac{-i}{2\pi^{3}}\sum_{n}\left[\frac{E_{n}X_{0n}^{\alpha}X_{n0}^{\beta}}{E_{n}-\hbar\omega}-\left(\omega\rightarrow-\omega\right)^{*}\right],
 \label{eq:conductivity}
\end{eqnarray}
where $\sigma_{0}=\frac{e^2}{4\hbar}$, an imaginary broadening is introduced via the transformation $\hbar\omega\rightarrow\hbar\omega+i\Gamma$, and the generic exciton dipole matrix element reads \cite{pedersen_intraband_2015,taghizadeh_nonlinear_2019,PhysRevB.107.235416}
\begin{eqnarray}
X_{0n}^{\alpha}=i\frac{\hbar}{m_{0}}\int\,\psi_{cv\mathbf{k}}^{\left(n\right)}\frac{p_{vc\mathbf{k}}^{\alpha}}{E_{c\mathbf{k}}-E_{v\mathbf{k}}}d^{2}\mathbf{k},
\end{eqnarray}
where $\alpha/\beta$ represent cartesian directions, $\omega$ is the frequency of the incident field, $E_n$ is the energy of the excitonic state $n$, $p_{vc\mathbf{k}}^{\alpha}$ is the interband momentum matrix element in the $\alpha$ direction, $m_0$ the bare electron mass, and $\psi_{cv\mathbf{k}}^{\left(n\right)}$ the $\mathbf{k}-$space exciton wave function for the state $n$, obtained by solving Bethe--Salpeter equation for the considered system. 

The interband momentum matrix elements for circularly polarized light can be written as a superposition of the matrix elements for linearly polarized light \cite{PhysRevB.52.14636,PhysRevB.99.235433,PhysRevB.107.235416} 
\begin{eqnarray}
p_{vc\mathbf{k}}^{\pm} & = & \frac{p_{vc\mathbf{k}}^{x}\pm i p_{vc\mathbf{k}}^{y}}{\sqrt{2}} \nonumber \\
&=&\frac{1}{\sqrt{2}}\left<v,\mathbf{k}\middle|\frac{m_{0}}{\hbar}\left[\frac{\partial\mathcal{H}\left(\mathbf{k}\right)}{\partial k_{x}}\pm i\frac{\partial\mathcal{H}\left(\mathbf{k}\right)}{\partial k_{y}}\right]\middle|c,\mathbf{k}\right>\nonumber \\
 \label{eq:p_circ}
\end{eqnarray}
where $\mathcal{H}$ is the tight--binding Hamiltonian and $m_0$ is the free-electron mass. This interband momentum matrix element includes both $p_{vc\mathbf{k}}^{x}$ and $p_{vc\mathbf{k}}^{y}$ components, where the $\pm$ sign defines left-handed (positive) and right-handed (negative) circularly polarized light \cite{PhysRevB.106.075403}.

Re-identifying the left- and right-handed polarization directions (LCP/RCP) with the $\pm$ sign, we can write the linear conductivity for circularly polarized light as
\begin{align*}
\frac{\sigma_{\pm}\left(\omega\right)}{\sigma_{0}}&=\frac{-i\hbar^{2}}{2\pi^{3}m_{0}^{2}}\sum_{n}\frac{E_{n}\left|X_{0n}^{\pm}\right|^{2}}{E_{n}-\hbar\omega}.\nonumber
\end{align*}
Directly substituting $p_{cv\mathbf{k}}^{\pm}$ from Eq. (\ref{eq:p_circ}) and separating the two terms, we have 
\begin{eqnarray}
    	X_{0n}^{\pm} & =i\int\,\psi_{cv\mathbf{k}}^{\left(n\right)}\frac{p_{vc\mathbf{k}}^{\pm}}{E_{c\mathbf{k}}-E_{v\mathbf{k}}}d^{2}\mathbf{k}\nonumber\\
	 & =\frac{1}{\sqrt{2}}\left[X_{0n}^{x}\pm iX_{0n}^{y}\right],
\end{eqnarray}
and then the linear conductivity for circular polarization is related to the linear polarization one by 
\begin{eqnarray}
    \sigma_{\mathrm{\pm}}\left(\omega\right) & =\frac{\sigma_{xx}\left(\omega\right)}{2}+\frac{\sigma_{yy}\left(\omega\right)}{2} \pm \frac{i}{2}\left[\sigma_{yx}\left(\omega\right)-\sigma_{xy}\left(\omega\right)\right].
\label{eq:circularconductivity}
\end{eqnarray}

Hence, we can compute the linear conductivity for circularly polarized light knowing only the various components from the optical response to linearly polarized light, having only to apply the Kramers--Kronig relations \cite{deL.Kronig:26,kramers1928diffusion,garfken67:math,10.1119/1.15901,jackson_classical_1999} to obtain the imaginary part of the off--diagonal components from their real part. Furthermore, the lattice symmetry of hBN in the nearest--neighbours tight--binding approach results in $\sigma_{yx}\left(\omega\right)=\sigma_{xy}\left(\omega\right)$, meaning that the linear conductivity for both circular polarization directions are equivalent.


\section*{SM4: Real-time simulations}
\label{sec:appendixD}

We use the EDUS code \cite{doi:10.1021/acs.jctc.2c00674} to compute the laser-driven dynamics in hBN. The numerical approach considers the light-matter and exciton interactions on the time domain. The approach evolves the one-electron reduced density matrix in reciprocal space. We use a grid of 200x200 points and periodic boundary conditions. The laser pump pulse that induces the exciton migration is modelled in time with a $\sin ^2$ envelope as
\begin{align}
	E(t)&=E_0 \sin \left(\omega\left(t-t_0\right)\right) \sin ^2\left(\pi \frac{t-t_0}{t_f-t_0}\right)\nonumber\\
    &=E_0 \sin \left(\omega t\right) \sin ^2\left(\pi \frac{t}{t_f}\right)\label{eq:pump-laser},
\end{align}
where $t_0$ is the beginning of the pulse, set at $t_0 =0$, $t_f=n_\mathrm{cycles}\frac{2\pi \hbar}{E_{\mathrm{avg}}}$ is the duration of the pulse, and $\omega$ is the photon frequency. The photon frequency is centered between the two 1s excitons peaks at the energy $\omega=E_{\mathrm{avg}}/\hbar$ in order excite both of them. The specific duration will depend on the average energy of the two $1s$ states as $T=n_\mathrm{cycles}\frac{2\pi \hbar}{E_{\mathrm{avg}}}$, where $n_\mathrm{cycles}=16$ is the number of cycles of the pulse, meaning the full duration of the pulse will be approximately $T\approx12.5\,\mathrm{fs}$ for the considered exciton states. The pulse intensity is set at $E_0=10^8\, \mathrm{W}\mathrm{cm}^{-2}$. In order to induce a superposition between the two non--degenerate $1s$ excitonic states, the pulse is left-handed circularly polarized.

For studying the core excitons, we also require a probe pulse. This new laser pulse is modeled in time with a Gaussian envelope as 
\begin{align}
    E(t)&=E_0 \sin \left(\omega^\prime \left(t- \tau_{\mathrm{d}}\right)\right) e^{-\frac{\left(t- \tau_{\mathrm{d}}\right)^2}{2 \sigma^2}},
\end{align}
where $ \tau_{\mathrm{d}}$ and $\sigma$ are the maximum and variance of the Gaussian envelope, respectively, and $\omega^\prime$ the photon frequency necessary to reach the core $1s$ orbital of Nitrogen $\left(\hbar\omega^\prime\approx409.9\,\mathrm{eV}\right)$ \cite{RevModPhys.39.78,thompson2001x,Malakhov2024}. To capture different time steps in the exciton migration, the maximum of the Gaussian envelope is delayed by different intervals $ \tau_{\mathrm{d}} = \Delta\tau_{\mathrm{d}} + T$, where $T$ is the duration of the $\sin^2$ pump pulse as defined previously and $\Delta\tau_{\mathrm{d}}$ is the delay between the two pulses. 

By the arrival of the probe pulse, the exciton state that is being probed is then given by
\begin{align}
    \left|\psi\left( \tau_{\mathrm{d}}\right)\right>&=e^{-i \frac{E_{+} \tau_{\mathrm{d}}}{\hbar}}\left[ \left|1s^{+}\right> + ie^{i\frac{\Delta E_{\mathrm{X}} \tau_{\mathrm{d}}}{\hbar}}\left|1s^{-}\right>\right].
\end{align}
The high-energy probe pulse then excites a core electron into a valence-band hole, which is left behind by the pump pulse interaction. Under the assumption of the sudden excitation, we expect the core exciton state to have the following form after the probe pulse
\begin{align}
    &\left|\psi_\mathrm{ce}\left(t^\prime\right)\right> = \sum_{\mathbf{k}}B_{\mathbf{k}}c^{\dagger}_{v\mathbf{k}}c_{\mathrm{co},\mathbf{k}}\left|\psi(\tau_d)\right> \\
    &=e^{-i \frac{E_{+} \tau_{\mathrm{d}}}{\hbar}}\sum_{\mathbf{k}}B_{\mathbf{k}}\left[c^{\dagger}_{v\mathbf{k}}c_{\mathrm{co},\mathbf{k}}\left|1s^+\right> + ie^{i\frac{\Delta E_{\mathrm{X}} \tau_{\mathrm{d}}}{\hbar}}c^{\dagger}_{v\mathbf{k}}c_{\mathrm{co},\mathbf{k}}\left|1s^-\right>\right]\nonumber
\end{align}
where $c_{co,\mathbf{k}}$ is the core annihilation operator. Core excitons are, therefore, defined over a new effective ground state which is the existing exciton, instead of the Fermi sea from which the initial exciton is defined. The coefficients $B_{\mathbf{k}}$ would be determined variationally as with regular excitons, and this would have to be done at each time $t'$.

Two distinct oscillations as a function of the pulse delay $\tau_{\mathrm{d}}$ are then present: firstly, the exciton migration that evolves as $e^{i\frac{\Delta E_{\mathrm{X}} \tau_{\mathrm{d}}}{\hbar}}$; 
secondly, the global phase of the excitonic state created by the pump pulse will go as $e^{-i \frac{E_{+} \tau_{\mathrm{d}}}{\hbar}}$. Those oscillations lead to distinct periods as a function of the time delay, namely 
\begin{align}
    T_\tau &= \frac{2\pi\hbar}{\Delta E_{\mathrm{X}}}, 
    &T_{\tau_+} &= \frac{2\pi\hbar}{E_+}.
\end{align}


The global phase gives rise to changes in the absorption line shape \cite{PhysRevResearch.3.013144,Picon_2019,science.1234407} with quite a fast period. While $T_\tau$ will be of the order of $10\,\mathrm{fs}$ for the considered strains, the much larger value of $E_+ \approx5.5\,\mathrm{eV}$ corresponds to $T_{\tau_+}\approx\,0.79\,\mathrm{fs}$. Such oscillations are presented in figure 3b by considering small steps in $\tau_{\mathrm{d}}$, and these are filtered from the panels c through e of figure 3 by considering steps in $\tau_{\mathrm{d}}$ which are multiples of $T_{\tau_+}$.


\bibliographystyle{ieeetr}
\bibliography{strain-tuning}